\documentclass{svjour3}                     
\smartqed  
\usepackage{graphicx}
\usepackage{fix-cm}
\journalname{Journal of Low Temperature Physics}

\begin{document}

\title{Diffusive Decay of the Vortex Tangle and Kolmogorov turbulence in
quantum fluids \thanks{This work was supported by
grants N 10-08-00369 and N 10-02-00514 from the RFBR.}}
\author{Sergey K. Nemirovskii \and L.P. Kondaurova }




\institute{Sergey K. Nemirovskii \at
             Institute of Thermophysics, Novosibirsk, Russia \\
            Tel.: +7-383-3309353\\
             \email{nemir@itp.nsc.ru}  \\         
         \and
      L.P. Kondaurova \at
          Institute of Thermophysics, Novosibirsk, Russia \\}

\date{Received: date / Accepted: date}
\maketitle
\begin{abstract}
The idea that chaotic set of quantum vortices can mimic classical
turbulence, or at least reproduce many main features, is currently actively being
developed. Appreciating
significance of the challenging problem of the classical turbulence it can
be expressed that the idea to study it in terms of quantized line is indeed
very important and may be regarded as a breakthrough. For this reason, this
theory should be carefully scrutinized. One of the basic arguments
supporting this point of view is the fact that vortex tangle decays at zero
temperature, when the apparent mechanism of dissipation (mutual friction) is
absent. Since the all possible mechanisms of dissipation of the vortex
energy, discussed in the literature, are related to the small scales, it is
natural to suggest that the Kolmogorov cascade takes the place with the flow of
the energy, just as in the classical turbulence. In a  series of recent experiment
attenuation of vortex line density was observed and authors attribute this decay
to the properties of the Kolmogorov turbulence. In the present work we discuss
alternative possibility of decay of the vortex tangle, which is not related
to dissipation at small scales. This mechanism is just the diffusive like
spreading of the vortex tangle. We discuss a number of the key experiments,
considering them both from the point of view of alternative explanation and
of the theory of Kolmogorov turbulence in quantum fluids.

\keywords{Superfluid Turbulence, Quantum vortices, Diffusive decay}
\PACS{PACS 67.25.dk, \and PACS 47.37.+q, \and 98.80.Cq }
\end{abstract}

\section{Introduction}


The diffusion-like  decay of the vortex tangle is closely related to the
hypothetical connection between the classical (Kolmogorov) and superfluid turbulence.
The idea that chaotic set of quantum vortices can mimic classical
turbulence, or at least reproduce many main features, is currently actively being
developed \cite{Brachet97},\cite{Skrbek99},\cite{Vinen00},\cite{Araki2002},%
\cite{Skrbek07},\cite{Golov08}. In principle, this idea had been discussed
early (see e.g. famous textbook by Frisch \cite{Frisch95}), as an alternative
version of the problem of classical turbulence. But only now, when the new
powerful experimental methods in quantum fluids appeared, this idea can be
checked experimentally and it seems to be very attractive. Appreciating
significance of the challenging problem of the classical turbulence it can
be expressed that the idea to study it in terms of quantized line is indeed
very important and may be regarded as a breakthrough. For this reason, this
theory should be carefully scrutinized.\\
    One of the basic arguments supporting the idea of Kolmogorov turbulence
in quantum fluids is the fact that
vortex tangle decays at zero temperature, when the apparent mechanism of
dissipation (mutual friction) is absent. Numerical and experimental
observations of decay of the tangle at zero temperature are presented in
papers \cite{Tsubota2000},\cite{mac-clintock},\cite{Pickett-2006},\cite%
{Golov07}. The physical mechanisms of the dissipation can be various, some
approaches and ideas such as a cascade-like break-down of the loops, Kelvin
waves cascade, acoustic radiation, reconnection loss, etc., have been
discussed in detail in recent review \cite{barenghi-decay}. It is remarkable
that all of these mechanisms are realized only on a very small scale.
Therefore, it is natural to suggest that the Kolmogorov cascade occurs with
the flow of energy, just as in the classical turbulence. The mentioned
experimental works on decay of the vortex tangle are interpreted from point
view of the decay of classical turbulence. In works by Skrbek \cite{Skrbek99}
it was developed approach, which relates the attenuation of the energy
in classical turbulence to  the decay  of the vortex line density and
predicts the temporal dependence of this attenuation.\\
    In the present work we discuss the alternative mechanism of  decay of the
vortex tangle, which is not related to dissipation at the small scales. This
mechanism is just the diffusive spreading of the vortex tangle. We applied
the diffusion equation obtained early (\cite{NemirPRB2010}) to describe
the decay of superfluid turbulence in the listed above experiment. Our
calculations enable us to conclude that mechanism of diffusion can describe
correctly the attenuation of the vortex line density. Besides, we discuss
these experiments from position of theory Kolmogorov turbulence.\\
    In parallel, to check our hypothesis, we perform direct numerical simulation
of the evolution of the vortex tangle, originally concentrated in the
restricted domain (See Fig.1). We were carefully monitoring all mechanisms
of the decrease of the total vortex length (reconnection procedure,
inserting or removing of intermediate points, elimination of small loops
etc.). It is found that the most effective mechanism is related with the
spreading of the vortex tangle, and the rate of change of the total length is
occurring in accordance with the diffusion equation, described above.

\begin{figure}[tbp]
\includegraphics[width=0.9\textwidth]{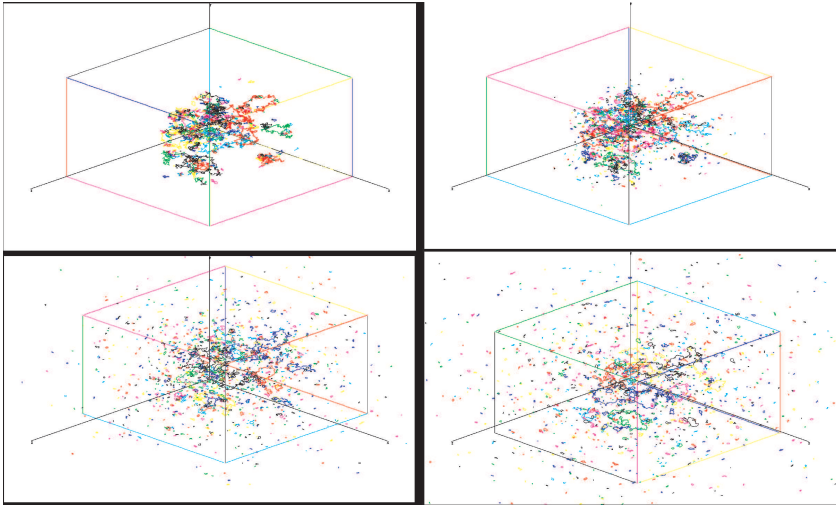}
\caption{Diffusion of a vortex tangle at different times. It is clearly seen
as the vortex loops escape from the volume carrying away the line
length and energy.}
\end{figure}

\section{Diffusion Equation}

In this section we very briefly describe main ideas leading to the diffusion
theory of the vortex loops, details can be found in paper \cite{NemirPRB2010}.
Vortex loops composing the vortex tangle can move as a whole with some drift
velocity $V_{l}$ depending on their structure and their length. The flux of
the line length, energy, momentum etc., executed by the moving vortex loops
takes place. In the case of inhomogeneous vortex tangle the net flux $%
\mathbf{J}$ of the vortex length due to the gradient of concentration of the
vortex line density $\mathcal{L}(x,t)$ appears. The situation here is
exactly the same as in classical kinetic theory with the difference being
that the "carriers" are not the point particles but the extended objects
(vortex loops), which possess an infinite number of degrees of freedom with
very involved dynamics. We offer to fulfill investigation basing on the
supposition that vortex loops have the Brownian or random walking structure
with the generalized Wiener distribution (see \cite{nemirPRB1998},\cite%
{NemirPRL2006},\cite{Nemir2008}).\\
To develop the theory of the transport
processes fulfilled by vortex loops (in spirit of classical kinetic theory)
we need to calculate the drift velocity $V_{l}$ and the free path $\lambda
(l)$\ for the loop of size $l$. Referring to the paper \cite{NemirPRB2010}
we write down here the following result. The drift velocity $V_{l}$ for the
loop of size $l$.is%
\begin{equation}
V_{l}=C_{v}\beta /\sqrt{l\xi _{0}},  \label{Vl}
\end{equation}%
Quantity $\beta $ is $(\kappa /4\pi )\ln (\mathcal{L}^{-1/2}/a_{0})$, where $%
\kappa $ is the quantum of circulation and $a_{0}$ is the core radius, $%
C_{v} $ is numerical factor of the order of unity. The $\xi _{0}$ is the
parameter of the generalized Wiener distribution, it is of order of the
interline space $\mathcal{L}^{-1/2}.$ The free path $\lambda (l)$\ \ for
loop of length $l$ is:%
\begin{equation}
\lambda (l)=1/2lb_{m}\mathcal{L}.  \label{freepath}
\end{equation}%
Here, $b_{m}$ is the numerical factor , approximately equal to $\
b_{m}\approx 0.2$ \ It is seen that free path $\lambda (l)$ is very small,
it implies only very small loops give a significant contribution to
transport processes.Knowing the averaged velocity $V_{l}$ of loops, and the
probability $P(x)$ (both quantities are $l$ -dependent), we can evaluate the
spatial flux of the vortex line density $\mathcal{L},$ executed by the
loops. The procedure is very close to the one in the classical kinetic
theory, with the difference being that the carriers have different sizes,
requiring additional integration over the loop lengths. Referring again to
paper \cite{NemirPRB2010} we write the flux $\mathbf{J}$ of vortex line is
proportional to $\nabla \mathcal{L}$ and, correspondingly, the
spatial-temporal evolution of quantity $\mathcal{L}$ obeys the diffusion
type equation%
\begin{equation}
\frac{\partial \mathcal{L}}{\partial t}=D_{v}\nabla ^{2}\mathcal{L},
\label{diffusion equation}
\end{equation}%
Our approach is a fairly   crude to claim a good quantitative description.
However, if we are to adopt the data grounded on the exact solution to the
Boltzmann type "kinetic" equation for vortex
loops distribution (\cite{Nemir2008}), we conclude that $%
C_{d}\approx 2.2$. \ Further we use the (\ref{diffusion equation}) to
describe the decay of superfluid turbulence in various experiments including
numerical simulations.

\section{Discussion of the experimental data and numerical simulations}

In this section we discuss several experiments and  numerical simulation
on the decay of the vortex,
which are usually considered as the ground for the Kolmogorov decay of the
superfluid turbulence. They are the so called Lancaster experiment\cite%
{Pickett-2006}, the Manchester experiment \cite{Golov07} and the Osaka
numerical simulation \cite{Tsubota2000}.

\subsection{Osaka numerical simulation}

Results of numerical simulation on the dynamics of the vortex tangle at zero
temperature made in Osaka \cite{Tsubota2000}, are frequently considered as a
base for conclusion that for zero temperature decay of the vortex tangle
occurs via various mechanisms realizing at small scales. Attenuation of the
vortex line density obtained in numerical simulation \cite{Tsubota2000}
is depicted in the upper picture of the Fig. 2.\\
\begin{figure}[tbp]
\includegraphics[width=8cm]{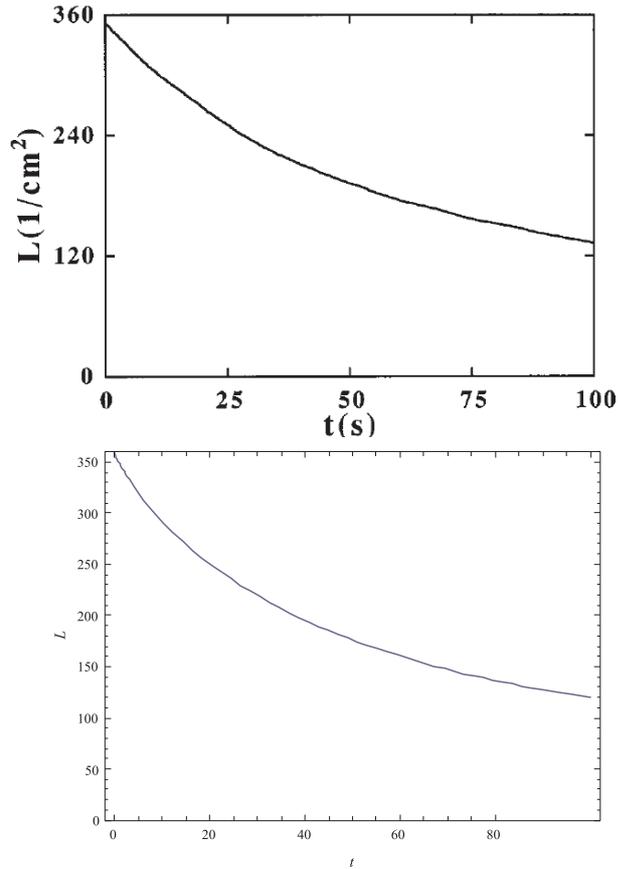}
\caption{Comparison with the numerical simulation by Tsubota et.al.%
\protect\cite{Tsubota2000} In the upper picture it is depicted the
attenuation of vortex line density obtained in numerical experiment
\protect\cite{Tsubota2000}. In the lower picture it is depicted the same
quantity calculated with the use of equation (\protect\ref{diffusion
equation}) without the auxiliary term, the diffusion constant was equal to $%
C_{d}\approx 2.2\ast 10^{-3}\ cm^{2}/s$. We calculated the two-dimensional
evolution of the vortex line density in the 1 cm square resolving
numerically equation (\protect\ref{diffusion equation}) with the boundary
conditions accounting the back radiation from the solid walls (see for details
\cite{NemirPRB2010} ).}
\end{figure}
Let us briefly
analyze the situation presented in paper \cite{Tsubota2000}, and demonstrate
that none of the currently discussed mechanisms of the "homogeneous" decay
of the vortex tangle at zero temperature, can be applied to this work.
Thus the Kelvin waves cascade could not be a reason for the vortex tangle
decay in numerical
simulation \cite{Tsubota2000}, simply because the space resolution $\Delta
\xi =2\ast 10^{-2}\ cm$ was too large to monitor the region of large wave
numbers, required for observation of a generation of higher harmonics.
Similarly, the acoustic radiation could not be a reason for the vortex tangle
decay, because the compressibility had not
been included in the numerical scheme. As for the loss of the line length
during reconnection, the real effect of the length loss can be obtained only on
the basis of more rigorous theory, e.g. with the use of the Gross-Pitaevskii
equation. It is known, however, that an artificial loss of length is
possible, due to realization of the reconnection procedure. This effect,
however, should depend on the space resolution, whereas it was proven that
the rate of decay did not depend on it. Feynman cascade of consequent breaking
down of vortex loops was imitated in \cite{Tsubota2000} with elimination
of very small loops (with sizes of few  $\Delta \xi $). But the tha total amount
of these events was is too
small to describe the decay of vortex line density, observed in the
numerical experiment.

To resume, one concludes that none of the discussed
mechanisms could be the reason of the "homogeneous decay" of the vortex
tangle in numerical simulation\cite{Tsubota2000}. Thus, the nature of
attenuation of the vortex line density in \cite{Tsubota2000} remained
unclear. The only mechanism capable of explain the decrease of the vortex
line density $\mathcal{L}(t)$ is just spreading of the tangle and escaping
of the vortex loops from the bulk of helium. To check our supposition, we
calculated the evolution of the vortex  line density under conditions
taking in work
\cite{Tsubota2000}resolving  equation (\ref%
{diffusion equation}). The problem had been resolved numerically, the
result is depicted in the lower picture of Fig.2. Comparison of the upper
and lower pictures of Fig.2, enabled us to conclude that the diffusion
spreading describes satisfactorily the evolution of the vortex tangle,
without any additional mechanism.

Resuming this subsection we can state that (i) there is no convincing
evidence in favor of existence of cascade-like transfer of the vortex length
(and energy)in direction of small scales and (ii)decay of the superfluid
turbulence is quite well described by a diffusion mechanism.

\subsection{Lancaster experiment}

Let us now discuss the recent experiment on decay of the vortex tangle at
very low temperatures \cite{Pickett-2006}. The authors of this
work reported the attenuation of vortex line density in
superfluid turbulent helium, $^{3}$He-B. In the upper picture of Fig.3, we
displayed Fig 2 of work \cite{Pickett-2006}, showing results of measurements
on the temporal behavior of the average vortex line density $\mathcal{L}(t),$
(solid curves, see \cite{Pickett-2006} for details).
\begin{figure}[tbp]
\includegraphics[width=7cm]{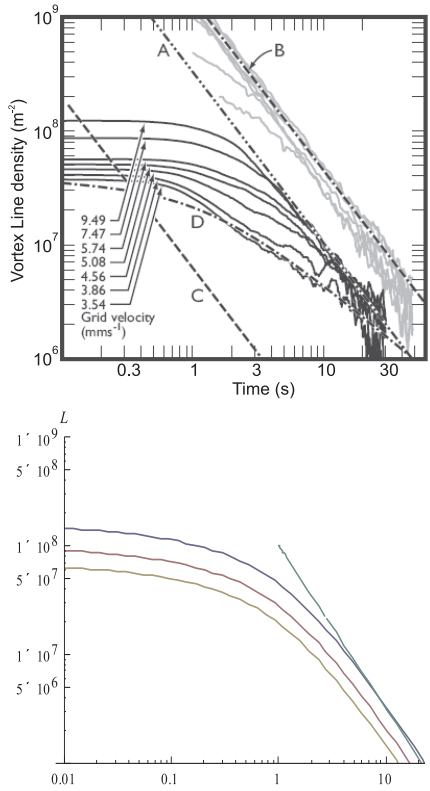}
\caption{Comparison with experiment \protect\cite{Pickett-2006}. See the
text Comparison with experiment \protect\cite{Pickett-2006}. In the upper
picture  we displayed Fig 2 of work \protect\cite{Pickett-2006},
showing results of measurements on the temporal behavior of the average
vortex line density $\mathcal{L}(t),$ (solid curves, see \protect\cite%
{Pickett-2006} for details). We calculated the same quantity resolving the
diffusion equation (\protect\ref{diffusion equation}), with the use the
boundary condition, which corresponds to the smearing of the vortex tangle
into ambient space. It is known that for this condition the solution of
three-dimensional problem is just production of three one-dimensional
solutions. The straight line in the lower picture exactly corresponds to line
A, in the upper picture (this line was named by the authors of paper
\protect\cite{Pickett-2006} as a "limiting behavior"). }
\end{figure}
Authors compare their
data with the theory by Skrbek \cite{Skrbek99} and conclude that decay of
the vortex tangle occurs in accordance with theory of classical (Kolmogorov)
turbulence. The main argument is that the long time attenuation of the
vortex length is described by the same line A ("limiting behavior") with the
$t^{-3/2}$ dependence (almost independently on initial level).
The according kinematic viscosity is $\nu ^{\prime
}\sim 0.2\kappa.$\\
It is necessary to note the following circumstance. The one of the standard
vision how the set of the vortex filament can imitate the classical turbulence
is that the lines are unified into the bundles (containing many lines). The
set of many bundles of various sizes randomly move, imitating the dynamics
of eddies in classical turbulence. Other view is that in the dense vortex
tangle there is polarization "indistinguishable by glance" and these polarized
vortices also reproduce the eddy dynamics.
It is necessary to note the following circumstance. The one of the standard
vision how the set of the vortex filament can imitate the classical
turbulence is that the lines are unified into the bundles (containing many
lines). The set of many bundles of various sizes randomly move, imitating
the dynamics of eddies in the classical turbulence. Other view is that in the
dense vortex tangle there is an averaged partial polarization of lines
"indistinguishable by glance" and
these polarized vortices also reproduce the eddy dynamics. Let us consider
the situation in the Lancaster experiment more attentively. Let us \ take some
"intermediate" value of the vortex line density $\mathcal{L}(t)=0.5\ast
10^{3}~cm^{-2}$, where all lines are collapsed into the "limiting", universal
behavior (line "A" in the upper picture of Fig.3). For this value the
interline space is equal to $\mathcal{L}^{-1/2}\approx \allowbreak 4.5\times 10^{-2}$
cm$.$ The latter implies that in the volume with size $d=1.5\ mm$ (offerred
by authors as the region, where the vortex tangle evolves) we have
approximately $d/L^{-1/2}\approx 3$ lines. Of course,  this amount is not
enough to form many bundles with very \textquotedblleft dense array of vortex
lines\textquotedblright , which are necessary to \textquotedblleft mimic
classical turbulence\textquotedblright . On the same ground one can assert
that it is hardly possible to say about the partial polarization of lines in the dense
tangle "indistinguishable by glance".\newline
Another fact is that about one third of the "limiting line A" is occupied
by the developed fluctuations, which
can blur the true dependence.\\
Consequently supposing that the decay is
realized by the escaping of the vortex loops we had estimated the
contribution into attenuation of the vortex line density, due to the pure
diffusion mechanism. We use the classical solution (see for details \cite%
{NemirPRB2010}). Results (with comments) are depicted in Fig. 3. We again
can  conclude that the diffusion
spreading describes satisfactorily the evolution of the vortex tangle,
without any additional mechanism.\\
Concluding this subsection we again can
state that (i) Interpretation of experiment is not fully consistent and cannot
serve as convincing evidence in favor of existence of
cascade-like transfer of the vortex length (and energy)in direction of small
scales, and (2)decay of the superfluid turbulence is quite well described by
the diffusion type mechanism.

\subsection{Manchester experiment}

The next experiment which we would like to discuss and which is also frequently
referred as the evidence of the Kolmogorov turbulence imitated by quanitized
vortex lines is the work \cite{Golov07}. In this work the decay of vortex
tangle in He-II was observed in the closed cube with solid walls. Results
are collected in the upper picture of Fig.4,where the temporal behavior of
the average vortex line density $\mathcal{L}_{av}(t)$ is depicted.
\begin{figure}[tbp]
\includegraphics[width=7cm]{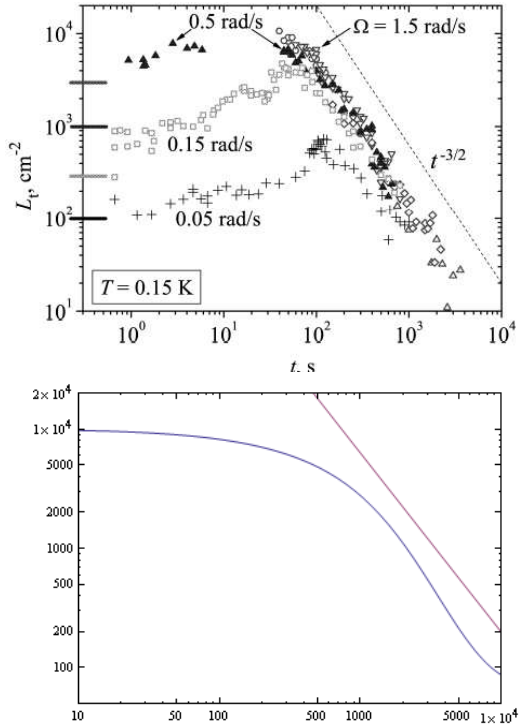}
\caption{Comparison with experiment \protect\cite{Golov07}. See the text
Comparison with experiment \protect\cite{Golov08}. In the left picture of
Fig. 6, it is depicted the temporal behavior of the average vortex line
density $\mathcal{L}_{av}(t)$ is depicted. We calculated the same dependence
on the basis of the diffusion equation (\protect\ref{diffusion equation}),
with the boundary boundary accounting back radiation of loops from the solid
walls. The full three dimensional problem had been resolved numerically.}
\end{figure}
Authors
compare their data with the theory by Skrbek \cite{Skrbek99} and conclude
that decay of the vortex tangle occurs in accordance with theory of
classical (Kolmogorov) turbulence. The according kinematic viscosity is $\nu
^{\prime }\sim 0.003\kappa $ , which about two orders smaller, than obtained
in the Lancaster experiment. Authors of \cite{Golov07} assumed that the
source of this discrepancy is that the turbulence observed in \cite%
{Pickett-2006} is not homogeneous, and the size of the energy-containing
eddies may differ from the spatial extent of the turbulence, so that the
value of $\nu ^{\prime }$ obtained in \cite{Pickett-2006} was ambiguous.\\
Coming to the previous subsection we can say that situation with the number
of vortices is better than in the Lancaster experiment. It is should be
noted however that it is merely due to larger size of the experimental cell. If
to take parts of experimental cell with sizes of the order 3 mm we
again meet the of very dilute vortex tangle. Another remark is related to the
Volovik observation \cite{Volovik03}. At the low temperature in $^{3}$He-B and
almost at any temperature in $^{4}$He, there should the so called Vinen
(not Kolmogorov)  turbulence, and, the whole
ideology grounded on the Skrbek theory is not applicable in the case of Manchester
experiment.

In our opinion the large differnce in the value of kinematic viscosity is $%
\nu ^{\prime }~$ in works \cite{Golov07} and \cite{Pickett-2006} (In fact it
is just difference in the total time of the decay) is related to (i) the
different sizes of the cells, where the superfluid turbulence is activated,
and (ii)
to the different boundary conditions (solid walls in \cite{Golov07}, and
absence of boundaries in \cite{Pickett-2006}). These two facts definitely
point out that the decay has the diffusive nature. We calculated the decay of
the vortex tangle on the basis of the diffusion equation (\ref{diffusion
equation}).  The fully three-dimensional problem was resolved numerically
(see for details  \cite{NemirPRB2010}), the result shown in the lower
picture of Fig. 4. It can be seen that the decay of the vortex tangle, due
to diffusion, describes both quantitatively and qualitatively the features
observed in the experiment. First of all, the whole qualitative behavior of lines agrees
with diffusive-like attenuation. In particular, there is a plateau, which is
changed with the fast decay of the tangle. Full decay of the superfluid
turbulence occurs in times, which are in a very good agreement, predicted on
the basis of the diffusion approach elaborated here. The slope of the curve
in the interval of the most intensive decrease, shows the dependence close
to $\sim t^{-3/2},$ which is also typical for diffusion.\\
Resuming this subsection we again claim that (i) experiment is not fully
self-consistent and (ii) the diffusion mechanism describes well the
attenuation of the vortex line density.

\subsection{Novosibirsk numerical simulation}

We fulfilled the direct numerical simulations of the evolution of the
nonuniform vortex tangle, originally concentrated in the restricted domain,
at zero temperature.The numerical simulation is performed with the use of
local induction approximation (LIA). An algorithm, which is based on
consideration of crossing lines, is used for vortex reconnection processes.
The calculations are carried out  in an infinite volume.  Result of the
numerical simulations is depicted in Fig.1, it is clearly seen
how the vortex loops escape from the volume and vortex line density
diminishes inside the bulk. We properly studied the change
of the total vortex length in the initial domain. In particular we were
carefully monitoring all mechanisms leading to the loss of the length, such
as the change of length owing to reconnection processes, the eliminations of
very small loos below the space resolution, the change of length due
insertion and removing of points to supply numerical algorithm stability and
etc. Results of this monitoring are depicted in Fig. 5. From this figure we
conclude, that the master mechanism of decrease of the vortex length inside
initial volume is related to the escaping of vortex loops, with is realized
in the diffusion like manner, and the other mechanisms listed above do not
affect appreciably on the spreading of the vortex tangle.\\
We compared results of our numerical simulations with the theory of
diffusion of the vortex tangle described above. We obtained, that the
evolution of of the length in initial domain is satisfactory described by
the with the diffusion equation (\ref{diffusion equation}). Result of
comparison is depicted in Fig 6.  The good
agreement with the experimental data and numerical simulation enables us to
conclude that the diffusion process plays a dominant role in the free decay
of the vortex tangle in the absence of the normal component.

\begin{figure}[tbp]
\includegraphics[width=0.6\textwidth]{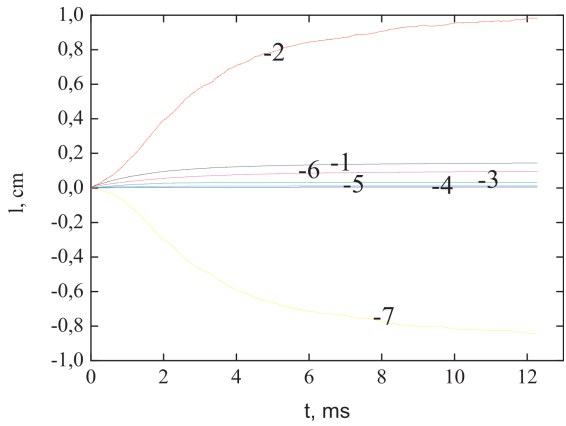}
\caption{Contribution of various mechanisms into decrease of total length.
1-Decrease of total length, 2-Decrease of total length inside domain
(spherical domain with the radius R< 0.008 cm) due to escaping of the small
loops, 3-Decrease of total length due to elimination of small loops (3 mesh
sizes ), 4-Change of total length due to artificial procedure -- inserting
or removing of intermediate points, 5-Change of total length due to
reconnection procedure, 6 -Change of total length due to motion after
reconnection, 7-Total length inside domain}
\end{figure}

\begin{figure}[tbp]
\includegraphics[width=0.9\textwidth]{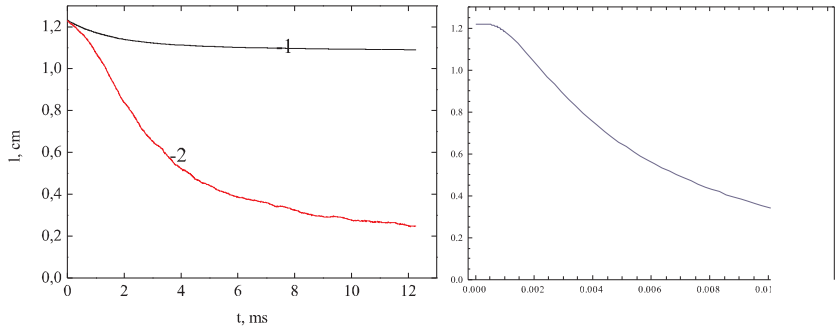}
\caption{Total length $l(t)=\protect\int \mathcal{L}(\mathbf{r,}t)d\mathbf{r}
$ ~inside domain obtained in numerical simulation (left picture) and
calculated on the base of the diffusion equation (for spherical domain)}
\end{figure}

\section{Conclusion}

Our main conclusion would be formulated as follows. The both experimental and
numerical data on decay of the superfluid turbulence discussed in the literature
cannot be regarded as the firm
evidence for the classical turbulence mechanism, accompanied by the Kolmogorov
cascade of the energy to region of very small scales. As we demonstrated, they
are not fully self-consistent even in the frame of the accepted approach. At
the time we had shown that all experimental data on decay of the vortex tangle
agree  well with the diffusion mechanism without any additional assumptions. Our
numerical results confirm this point of view, demonstrating that all possible
mechanisms of the loss of the vortex length (and,correspondingly, the vortex
energy)are considerably less than the loss of energy due to escaping of
the loops from the volume.
%



\begin{acknowledgements}
Authors are grateful to participants of workshop "SYMPOSIA ON SUPERFLUIDS
UNDER ROTATION" for the very useful discussion.
\end{acknowledgements}


\end{document}